\documentclass[%
 reprint,
 amsmath,amssymb,
 aps,
]{revtex4-2}

\usepackage{graphicx}
\usepackage{dcolumn}
\usepackage{bm}
\usepackage{hyperref}
\usepackage{siunitx}

\begin{document}

\title{Precision cross section measurements of neutron-induced non-elastic gamma production reactions at 14~MeV}

\author{Mauricio Ayllon Unzueta}
\affiliation{Lawrence Berkeley National Laboratory, 1 Cyclotron Road, Berkeley, CA 94720, USA}

\author{Emanuel Chimanski}
\affiliation{Brookhaven National Laboratory, 98 Rochester St, Upton, NY 11973, USA}

\author{Juan Cristhian Luque Gutierrez}
\affiliation{Lawrence Berkeley National Laboratory, 1 Cyclotron Road, Berkeley, CA 94720, USA}
\affiliation{North Carolina State University, 2500 Stinson Drive, Raleigh, NC 27695, USA}

\author{Ann M. Parsons}
\affiliation{NASA Goddard Space Flight Center, 8800 Greenbelt Rd, Greenbelt, MD 20771, USA}

\author{Patrick N. Peplowski}
\affiliation{The Johns Hopkins Applied Physics Laboratory, 11101 Johns Hopkins Road, Laurel, MD 20723, USA}

\author{Arun Persaud}
\affiliation{Lawrence Berkeley National Laboratory, 1 Cyclotron Road, Berkeley, CA 94720, USA}

\author{Jack T. Wilson}
\affiliation{The Johns Hopkins Applied Physics Laboratory, 11101 Johns Hopkins Road, Laurel, MD 20723, USA}

\date{\today}

\begin{abstract}
We present a technique for high-precision absolute measurements of $\gamma$-ray production cross sections $(n,x\gamma)$ induced by 14~MeV neutrons. The technique is based on the Associated Particle Imaging (API) method, which tags individual neutrons emitted from a deuterium--tritium source with their associated $\alpha$-particles, enabling coincidence-based suppression of lower-energy neutrons and room background signals, while also providing neutron flux measurements with uncertainties on the order of 1\%.

This compact, laboratory-scale technique has the potential to address gaps and discrepancies in existing cross section libraries at a fraction of the cost of large-scale dedicated facilities, with direct applications in active neutron interrogation, detector calibration, nuclear fusion science, and Monte Carlo simulation codes, among others. We demonstrate the technique through proof-of-concept experiments on thin and thick samples of natural Fe (1~mm, 8~mm) and natural C (2~mm, 10~mm).

For Fe, $\gamma$-ray production cross sections were measured for the 846.78 and 1238.33~keV transitions from the first and second excited states of $^{56}$Fe. For C, the $(n,n'\gamma)$ cross section was measured for the 4438.91~keV transition from the first excited state of $^{12}$C. Measurements were performed at $110^\circ$ and $48^\circ$ relative to the neutron beam to characterize $\gamma$-ray anisotropy.

For the thin samples, the measured cross sections are $727 \pm 71$~mb ($^{56}$Fe, 846.78~keV, $110^\circ$), $304 \pm 42$~mb ($^{56}$Fe, 1238.33~keV, $110^\circ$), $759 \pm 106$~mb ($^{56}$Fe, 846.78~keV, $48^\circ$), $321 \pm 151$~mb ($^{56}$Fe, 1238.33~keV, $48^\circ$), $141 \pm 16$~mb ($^{12}$C, 4438.91~keV, $110^\circ$), and $279 \pm 105$~mb ($^{12}$C, 4438.91~keV, $48^\circ$). Uncertainties are dominated by counting statistics and detector efficiency calibration, both of which can be reduced in future experiments to achieve overall uncertainties of 5\% or better.
\end{abstract}

\maketitle

\section{\label{sec:intro}Introduction and background}

Neutron-induced reactions followed by $\gamma$-ray emission, such as non-elastic scattering $(n,x\gamma)$ and radiative capture $(n,\gamma)$, enable nondestructive elemental assays in applications ranging from nuclear material characterization and contraband detection to oil-well logging and planetary science, among many others~\cite{Valkovic2016,Bishnoi2020,Carasco2008,Perot2009,Bishnoi2020a,Pozzi2023,Radtke2012,Prettyman2007}. These techniques rely on interpreting $\gamma$-ray emissions from neutron-irradiated samples, a process that depends heavily on radiation transport simulations and the underlying nuclear data libraries.

However, the accuracy of such assays is limited by inconsistencies in existing nuclear data. Studies in oil-well logging~\cite{Mauborgne2020} and planetary science (e.g.,~\cite{Yamashita2013,Prettyman2006,Kim2012,Lim2017}) have identified discrepancies between measured and simulated spectra, often traced to deficiencies in cross section libraries. Recent community-led nuclear data workshops~\cite{Kolos2022} have echoed these concerns, underscoring the need for improved experimental benchmarks to refine nuclear data libraries and enhance the reliability of simulation tools.

Deuterium-tritium (DT) neutron sources are commonplace in active interrogation applications, making precise cross sections at 14~MeV incident neutron energy essential for proper assay characterization. Current methods, however, are limited primarily by the accuracy of neutron flux measurements on the sample. The prevailing standard, activation foils irradiated alongside the sample, using well-characterized reference reactions e.g., $^{63}$Cu$(n,2n)^{62}$Cu, $^{65}$Cu$(n,2n)^{64}$Cu, $^{63}$Cu$(n,\alpha)^{60}$Co, $^{27}$Al$(n,\alpha)^{24}$Na, $^{58}$Ni$(n,p)^{58}$Co, and $^{93}$Nb$(n,2n)^{92m}$Nb, imposes a hard lower bound on the achievable uncertainty equal to that of the reference cross section itself.

We developed a technique that overcomes this limitation, providing high-precision neutron flux measurements as well as strong background suppression. The TANGRA collaboration has pursued related work~\cite{Prusachenko2025,Prusachenko2025a}, reflecting the growing community interest in these kinds of measurements. Our approach is nevertheless distinct in two key aspects: first, we employ small and thin samples rather than the large, thick samples used by TANGRA (e.g., Fe slabs of ${\sim}44 \times 44 \times 0.9$~cm), hence our geometry favors single-neutron scattering physics; secondly, we determine the neutron flux by imaging the $\gamma$-ray producing target directly, rather than mapping the tagged neutron beam profile with ancillary neutron detectors, enabling a more precise flux determination.

\section{\label{sec:methods}Methods}

\subsection{\label{sec:api}Associated Particle Imaging for $(n,x\gamma)$ cross section measurements}

DT neutron generators are compact devices that produce nearly monoenergetic, isotropic 14~MeV neutrons by accelerating a narrow beam of deuterium (D) and tritium (T) ions onto a titanium target, where they accumulate and undergo fusion reactions:
\begin{equation}
\label{eq:dt}
\text{D} + \text{T} \rightarrow n\,(14.1\;\text{MeV}) + \alpha\,(3.5\;\text{MeV}).
\end{equation}
This reaction emits a neutron ($n$) and an $\alpha$-particle in opposite directions in the center-of-mass frame due to conservation of momentum.

\begin{figure}[tb]
\includegraphics[width=\columnwidth]{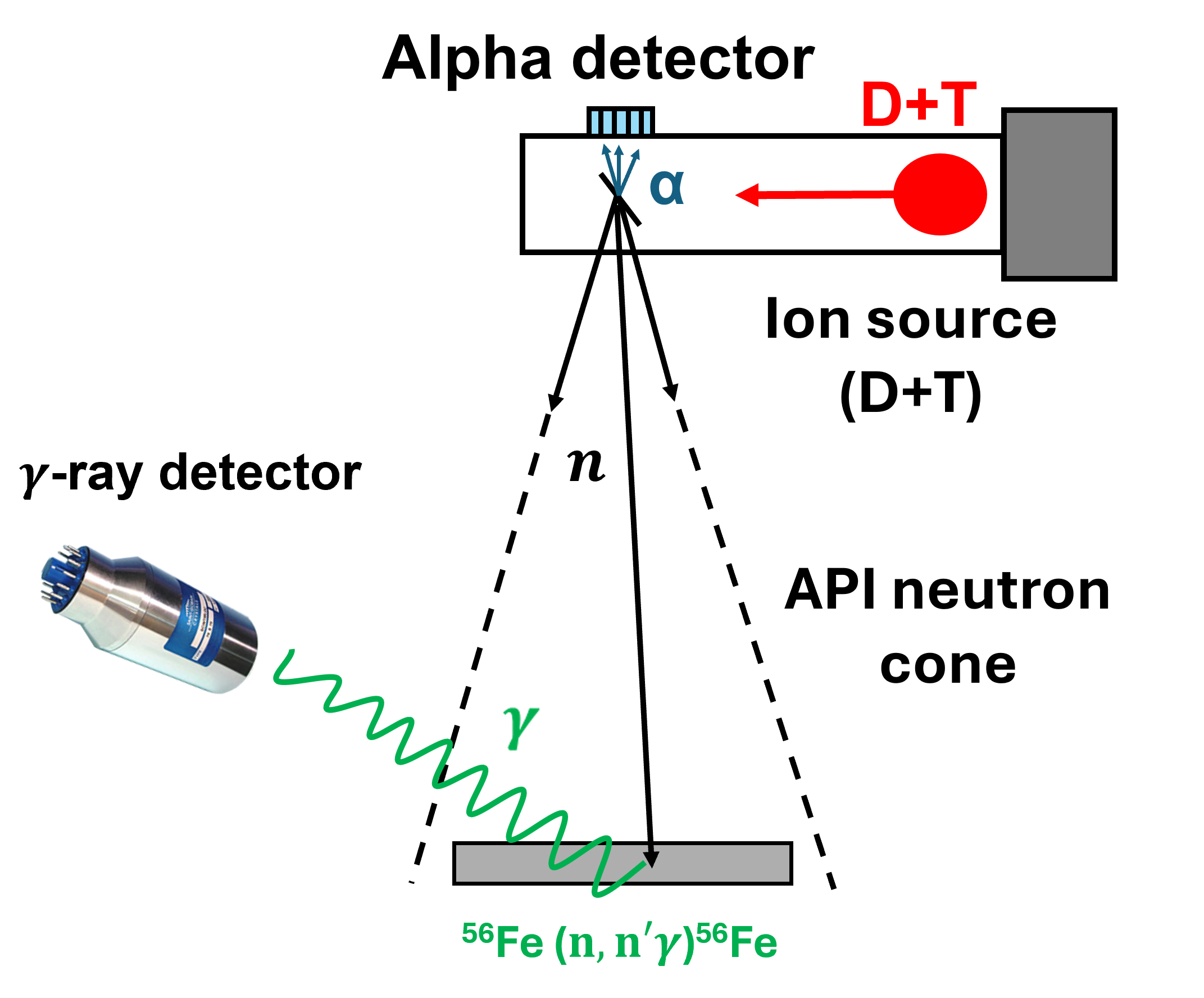}
\caption{\label{fig:api}Schematic of an API system showing a tagged neutron inducing prompt $\gamma$-ray emission in a thin iron sample.}
\end{figure}

Associated Particle Imaging (API) is a nuclear imaging technique that leverages this simultaneous back-to-back emission to obtain the start time and direction of each individual neutron, enabling three-dimensional (3D) compositional reconstruction. This neutron ``tagging'' is accomplished by integrating a position-sensitive $\alpha$-particle detector into the neutron generator and recording $\alpha$-particles in coincidence with neutron-induced $\gamma$-rays, as shown in Fig.~\ref{fig:api}. By inducing DT fusion in a precisely defined area with a small beamspot, the position of each $\alpha$-particle in the $\alpha$-particle detector can be used to reconstruct the 3D velocity vector of the associated neutron, while its arrival time provides the exact moment the DT reaction occurred. Coincident detection of the $\alpha$-particle and either a scattered neutron or a neutron-induced $\gamma$-ray then enables time-of-flight (TOF) measurements, providing the 3D coordinates of each scattering center in the sample. Previous studies have shown that centimeter-scale position resolution can be achieved with this API system~\cite{AyllonUnzueta2021}.

This approach offers several key advantages for cross section measurements:
\begin{itemize}
\item \textit{Precise neutron yield determination}: Every neutron directed toward the sample has a corresponding detected $\alpha$-particle. This enables a direct, low-uncertainty measurement of the neutron flux on sample, which is the dominant source of uncertainty in conventional methods that rely on calibrated detectors or co-irradiated activation foils.
\item \textit{Suppression of lower-energy reactions}: The coincidence requirement, with a time window of tens of nanoseconds, ensures that only prompt neutron-induced reactions are recorded, effectively suppressing contributions from lower-energy neutrons.
\item \textit{Room background suppression}: The 3D imaging capability enables spatial gating around the sample volume, significantly reducing background from surrounding materials.
\item \textit{Well-constrained neutron energy}: The nearly monoenergetic nature of the DT reaction, combined with knowledge of the sample's solid angle relative to the generator, tightly constrains the incident neutron energy at around 14~MeV.
\end{itemize}

Fig.~\ref{fig:api} illustrates a tagged neutron interaction with a $^{56}$Fe nucleus via inelastic scattering, followed by prompt $\gamma$-ray emission. By counting these $\gamma$-rays over a measurement period, together with the total number of detected $\alpha$-particles and the detector efficiency, the reaction cross section can be extracted directly. For the experiments described below, we used a LaBr$_3$ and a CeBr$_3$ $\gamma$-ray detectors, a Yttrium Aluminum Perovskite (YAP:Ce) $\alpha$-particle detector~\cite{CRYTUR}, an Adelphi DT108 neutron generator~\cite{Adelphi}, a PIXIE16 data acquisition system (DAQ)~\cite{XIA}, and a custom-built $\alpha$-particle detector assembly with associated electronics~\cite{AyllonUnzueta2019}. A complete characterization of the API system can be found in Ref.~\cite{AyllonUnzueta2021}.

\subsection{\label{sec:xsec-eq}Cross section equation and observables}

The cross section for 14~MeV neutron-induced $\gamma$-ray production is derived from the standard reaction rate equation:
\begin{equation}
\label{eq:reaction-rate}
R = N_t \, \sigma \, \phi
\end{equation}
where $R$ is the number of prompt reactions per second, $N_t$ is the number of sample atoms corrected for isotopic abundance, $\sigma$ is the $\gamma$-ray production cross section (barns), and $\phi$ is the neutron flux (n/cm$^2$/s) incident on the sample. $N_t$ is obtained from the measured sample mass ($m$), Avogadro's number ($N_A$), molar mass ($M$), and isotopic abundance ($f$):
\begin{equation}
\label{eq:Nt}
N_t = \frac{m \, N_A \, f}{M}
\end{equation}

In practice, we do not observe $R$ directly. Instead, we measure the net rate of detected $\gamma$-rays under a specific photopeak (e.g., 847~keV), which is related to the reaction rate by:
\begin{equation}
\label{eq:Rgamma}
R_\gamma = R \,  F \, \epsilon
\end{equation}
where $F$ accounts for $\gamma$-ray self-shielding through the sample and $\epsilon$ is the absolute detector efficiency (geometric and intrinsic). Combining Eqs.~(\ref{eq:reaction-rate}) and~(\ref{eq:Rgamma}) and solving for $\sigma$:
\begin{equation}
\label{eq:sigma-rate}
\sigma = \frac{R_\gamma}{N_t \, \phi \, F \, \epsilon}
\end{equation}

We can eliminate the time dependence of Eq.~(\ref{eq:sigma-rate}) by reformulating it in terms of the net counts under the photopeak ($N_\gamma$) and the time-integrated flux, or neutron fluence ($\Phi$):
\begin{equation}
\label{eq:sigma}
\sigma = \frac{N_\gamma}{N_t \, \Phi \, F \, \epsilon}
\end{equation}

This formulation is particularly convenient for API data analysis, given that we can directly measure the absolute number of tagged $\alpha$-particles with corresponding neutrons directed toward the sample, rather than relying on a time-averaged quantity. This approach naturally accounts for small variations in neutron yield over the course of the measurement and represents yet another advantage of the API technique for cross section measurements.

The cross section is, in general, a function of both neutron energy and angle (measured with respect to the incident neutron direction). However, the neutron energy is approximately constant at 14~MeV (see the discussion section). The differential form of Eq.~(\ref{eq:sigma}) is then as follows:
\begin{equation}
\label{eq:sigma-diff}
\frac{d\sigma}{d\Omega}(\theta) = \frac{N_\gamma(\theta)}{N_t \, \Phi \, F(\theta) \, \epsilon(\theta)}
\end{equation}

The determination of $N_\gamma(\theta)$ and $N_t$ is straightforward, but the neutron fluence and $F(\theta)\epsilon(\theta)$ warrant more detailed treatment, as described in the following subsections.

\subsection{\label{sec:fluence}Neutron fluence determination}

The data acquisition system (DAQ) includes an independent high-rate counter that records both coincident and noncoincident $\alpha$-particles via a fast trigger. This total number of detected $\alpha$-particles provides a direct proxy for the number of neutrons emitted into the API cone. Given the fraction of these neutrons directed toward the sample, the neutron fluence on target can be determined.

\begin{figure}[tb]
\includegraphics[width=\columnwidth]{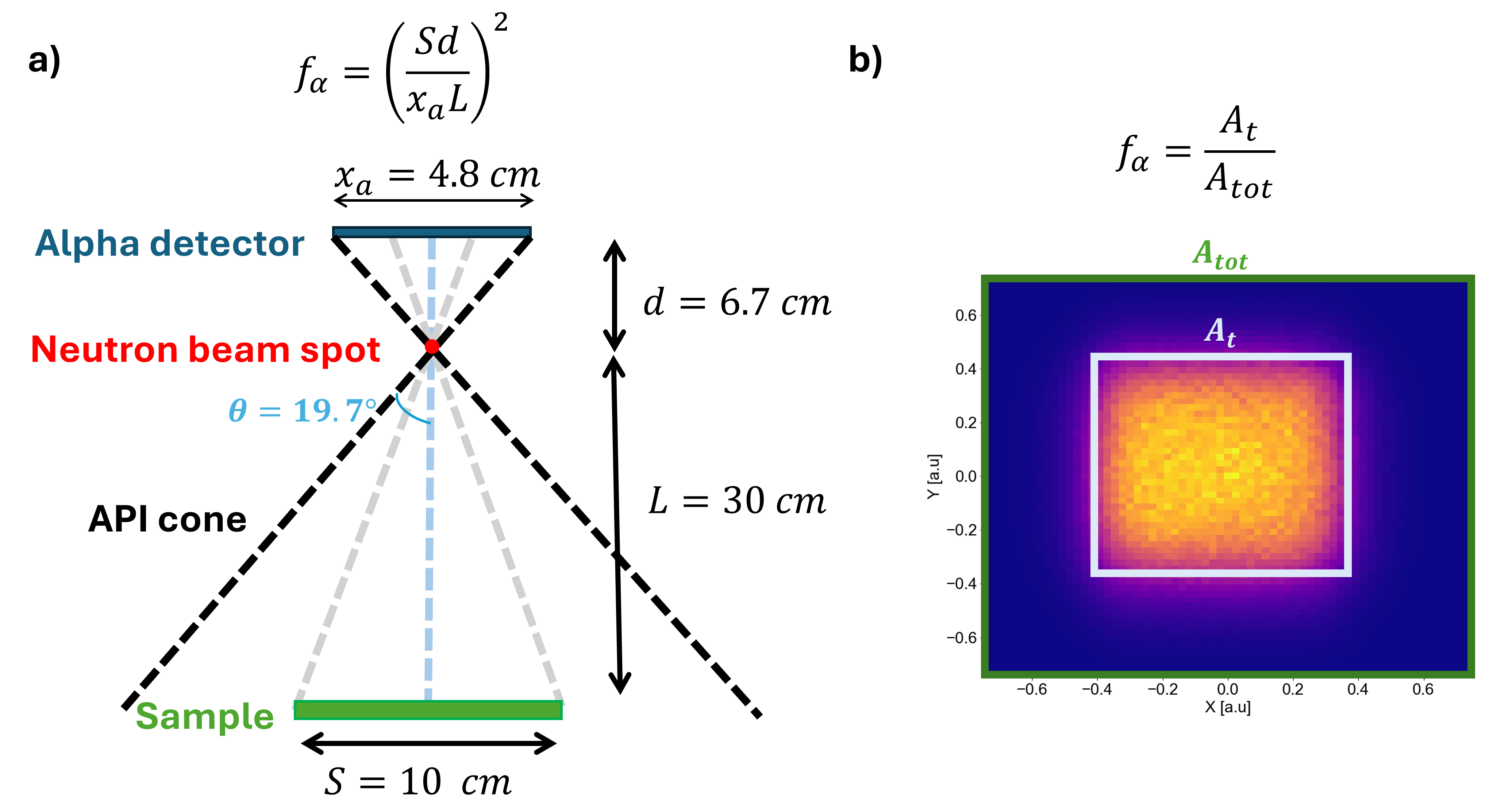}
\caption{\label{fig:fa}Two different ways of calculating $f_\alpha$, the fraction of associated $\alpha$-particles with corresponding neutrons directed into the sample: (a) from geometric considerations and (b) from area ratios obtained via 2D image projections onto the $\alpha$-particle detector.}
\end{figure}

Fig.~\ref{fig:fa}a shows a schematic of the experimental setup with the geometrical parameters relevant to deriving the neutron flux on the sample.

We define $f_\alpha$ as the fraction of detected $\alpha$-particles whose associated neutrons are directed into the sample. Geometrically, $f_\alpha$ can be thought of as the ratio of the sample area to the area of the API neutron cone at the height of the sample, projected back onto the $\alpha$-particle detector, as illustrated by the grey dashed lines in Fig.~\ref{fig:fa}a. This gives:
\begin{equation}
\label{eq:fa-geom}
f_\alpha = \left(\frac{S \, d}{x_\alpha \, L}\right)^2
\end{equation}
where $S$ is the side length of the square sample, $d$ is the distance between the center of the $\alpha$-particle detector and the neutron beam spot, $x_\alpha$ is the side length of the active region of the $\alpha$-particle detector, and $L$ is the distance from the neutron beam spot to the center of the sample. The neutron fluence is then:
\begin{equation}
\label{eq:fluence}
\Phi = \frac{N_\alpha \, f_\alpha}{A_t}
\end{equation}
where $N_\alpha$ is the total number of detected $\alpha$-particles (recorded directly by the fast counter in the DAQ) and $A_t$ is the sample area ($10\times10$~cm$^2$ in our setup).

While $f_\alpha$ can, in principle, be computed purely from geometrical considerations via Eq.~(\ref{eq:fa-geom}), this requires precise knowledge of parameters $d$ and $x_\alpha$, which are internal to the sealed neutron generator and therefore not accessible for direct measurement, since the generator is sealed to prevent tritium contamination. In addition, the beam spot size contributes to $f_\alpha$ but has not been independently characterized, both because the generator is sealed and also because it depends on operating conditions such as acceleration voltage and plasma pressure. Small uncertainties in any of these parameters can contribute significantly to the overall flux uncertainty, as discussed further in the Results section.

To avoid these limitations, we adopt an alternative approach that determines $f_\alpha$ directly from the projected sample image on the $\alpha$-particle detector (Fig.~\ref{fig:fa}b), requiring no knowledge of internal generator dimensions or the beam spot size. In the ideal case of perfect spatial resolution, $f_\alpha$ would simply equal the ratio of the projected sample area $A_t$ to the total sensitive detector area $A_\text{tot}$. In practice, finite position resolution prevents a hard geometric cut, so we instead apply a soft cut using two measurements: (1) a flat-field (FF) map, obtained from a 2D non-coincident measurement of associated $\alpha$-particles, and (2) a high-statistics coincident measurement of the square sample. A 2D function $g(x,y)$ is then fitted to the sample image and superimposed on the flat-field, yielding:
\begin{equation}
\label{eq:fa-image}
f_\alpha = \frac{A_t}{A_{tot}}=\frac{\int \int g(x,y) \, \text{FF} \, dx \, dy}{\int \int \text{FF}(x,y) \, dx \, dy}
\end{equation}

The neutron fluence $\Phi$ is then calculated from Eq.~(\ref{eq:fluence}).

\subsection{\label{sec:efficiency}Absolute efficiency and self-shielding}

The product $F \epsilon$ captures both the $\gamma$-ray self-shielding through the sample and the absolute detector efficiency, respectively. Both are functions of $\gamma$-ray energy and position, and they are determined together via Monte Carlo (MC) simulations using GEANT4 v11~\cite{Agostinelli2003}, whose accuracy was benchmarked with detailed detector characterization measurements. That process is detailed in Appendix~A of Ref.~\cite{Peplowski2026}, which includes the characterization of the same CeBr$_3$ detector used for this study. In summary, a series of measurements in the ANSI N42.14 geometry using well-characterized $\gamma$-ray source standards is used to develop a high-fidelity model of the detector geometry within the MC model. The model is adjusted as necessary to reproduce the source standard measurements. The final geometry reproduces the source measurements to within ${\sim}8.5\%$. Note that the systematic uncertainty on the source activity is 3\%, as provided by the manufacturer. Using different check sources with lower uncertainties will allow us to reduce the overall cross section uncertainty in future measurements.

The validated MC model is then used to determine $F \epsilon$ for the cross section experiment. This is done by placing the detector model and sample within the MC world volume in a geometry that is identical to the experiment. For each $\gamma$-ray energy of interest, a separate simulation is performed. For a given energy, $10^7$ unique $\gamma$-ray histories are simulated. For each event, $\gamma$-rays are generated uniformly within the sample volume, with an isotropic angular emission distribution. An energy-deposition histogram is generated for the simulation, and the number of events in the photopeak is divided by the number of histories to produce the $F \epsilon$ product. Because the simulation accounts for scattering within the sample and detector housing, the detector-to-sample geometry, and the $\gamma$-ray detection physics within the detector volume, the $F \epsilon$ product accounts for all intrinsic and geometric effects that impact $\gamma$-ray detection.

\section{\label{sec:setup}Experiment setup}

\begin{figure}[tb]
\includegraphics[width=\columnwidth]{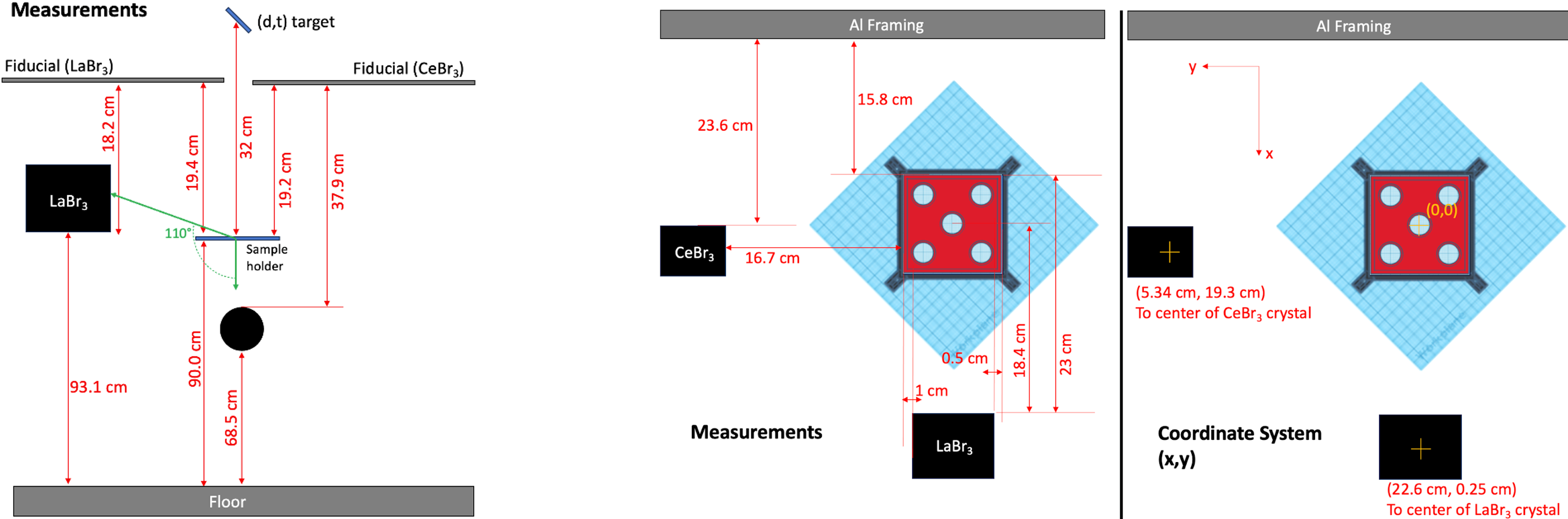}
\caption{\label{fig:setup}Schematic of the sample (Fe or C) relative to the neutron source and $\gamma$-ray detectors. The five circular holes in the sample holder indicate the positions of the standard check sources used for efficiency calibration.}
\end{figure}

The geometry of the experimental setup, including all relevant dimensions and the target-holder arrangement, is shown in Fig.~\ref{fig:setup}. We performed four sets of measurements using two high-purity natural iron samples (1~mm and 8~mm thick) and two natural carbon samples (2~mm and 10~mm thick). All samples have a $10 \times 10$~cm$^2$ cross-sectional area perpendicular to the API neutron beam and were positioned at the same location across all runs.

The neutron source is located 32~cm above the sample. The front face of the LaBr$_3$ crystal is positioned 18.4~cm from the center of the sample at an angle of approximately $110^\circ$ with respect to the neutron beam direction, while the front face of the CeBr$_3$ crystal is located 21.7~cm from the sample center at approximately $48^\circ$. Each sample is held in place by a low-density 3D-printed plastic holder (13.9~g of PLA) suspended by four Kevlar strings, minimizing the surrounding mass and thereby reducing background signals by limiting the amount of non-sample material within the $xy$ and $dt$ cuts to the data. Accounting for the solid angle subtended by each detector, the effective angular coverage is $110^\circ \pm 7^\circ$ for LaBr$_3$ and $48^\circ \pm 5^\circ$ for CeBr$_3$ with respect to the neutron beam direction, as obtained from a Monte Carlo simulation. These parameters are summarized in Table~\ref{tab:geometry}.

\begin{table}[b]
\caption{\label{tab:geometry}$\gamma$-ray detector properties and configuration.}
\begin{ruledtabular}
\begin{tabular}{lccc}
Detector & Size & Location $(x,y,z)$\footnote{Location of the center of the face of the detector with respect to the center of the sample.} & Angle\footnote{Angle with respect to the neutron beam direction.} \\
\colrule
LaBr$_3$ & 3'' & $(18.4, 0.25, 5.01)$~cm & $110^\circ \pm 7^\circ$ \\
CeBr$_3$ & 2'' & $(2.5, 21.7, -21.25)$~cm & $48^\circ \pm 5^\circ$ \\
\end{tabular}
\end{ruledtabular}
\end{table}

\section{\label{sec:results}Results}

In the following subsections, we present results for the four irradiated samples: thin and thick Fe, and thin and thick C. The results are organized according to the individual terms of Eq.~(\ref{eq:sigma}): the neutron fluence ($\Phi$) or flux, the net photopeak counts ($N_\gamma$), and the combined detector efficiency and sample self-attenuation factor ($\epsilon F$). The section concludes with the extracted prompt $\gamma$-ray production cross sections and their associated uncertainties.

\subsection{\label{sec:flux}Neutron flux}

\begin{figure}[tb]
\includegraphics[width=\columnwidth]{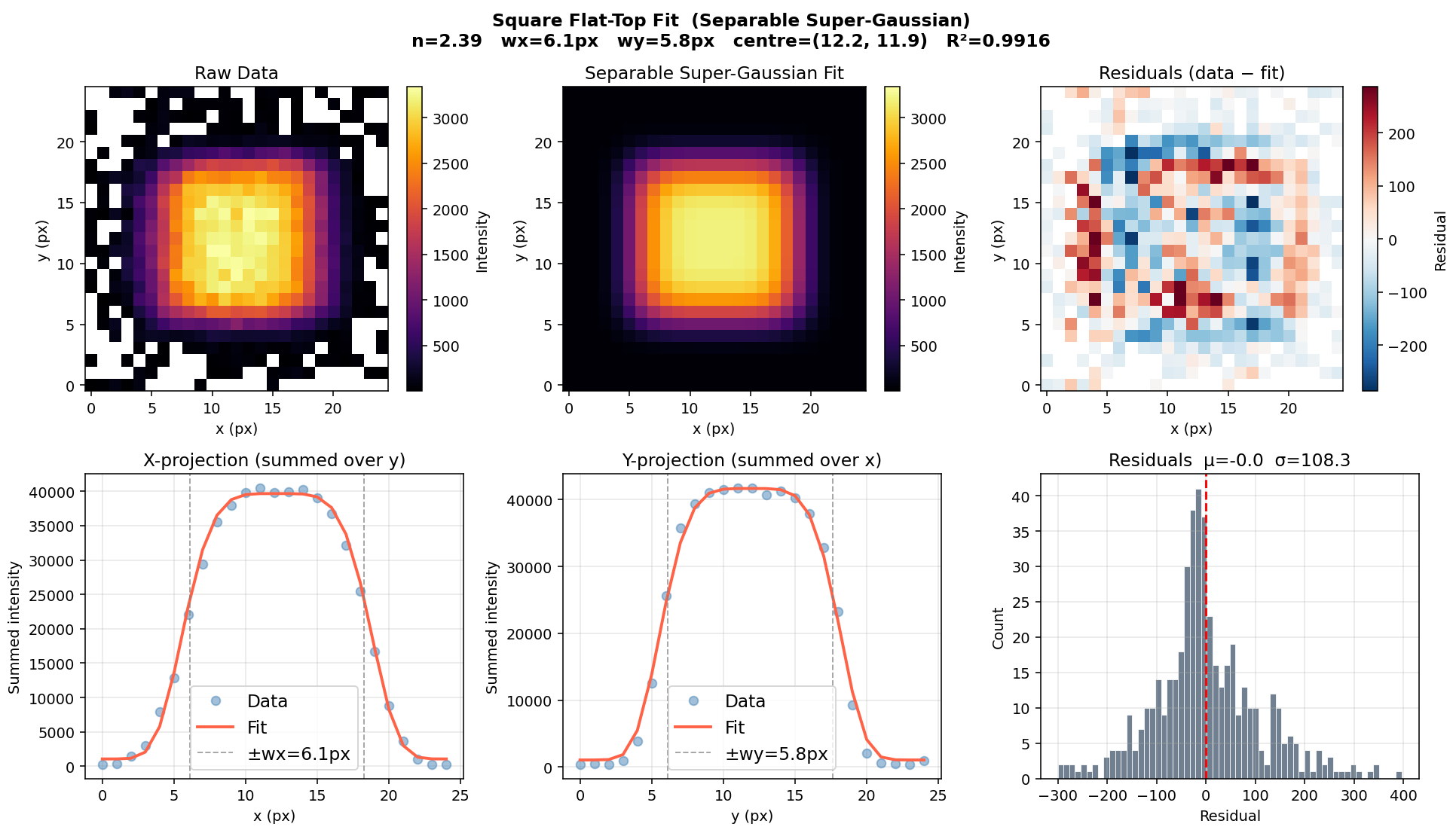}
\caption{\label{fig:fitting}An 8~mm-thick iron sample 2D profile as imaged by the API system and the results of the fitting algorithm used to obtain the function $g(x,y)$.}
\end{figure}

The parameter $f_\alpha$ was determined using measurements made with an 8-mm thick Fe sample that was irradiated at 65~kV for 4~hours at an estimated neutron yield of $5.5 \times 10^6$~n/s. Fig.~\ref{fig:fitting} shows some of the details of the fitting procedure used to obtain $g(x,y)$ in Eq.~(\ref{eq:fa-image}).

The raw 2D image, which represents the position of each $\alpha$-particle in an alpha-gamma coincidence with the LaBr$_3$ detector, was corrected by $1/r^2$, where $r$ is the distance from the corresponding sample position and the LaBr$_3$ detector. A separable super-Gaussian (square flat-top with Gaussian edges) was used for the fit:
\begin{equation}
\label{eq:supergauss}
g(x,y) = A \exp\!\left[-\left(\frac{x}{w_x}\right)^{2n} - \left(\frac{y}{w_y}\right)^{2n}\right] + B
\end{equation}
where $A$ is the amplitude, $w_x$ and $w_y$ are the half-width parameters, $n$ is the shape parameter, and $B$ is a constant background. The fitting routine provides uncertainties on all parameters, which are propagated through Monte Carlo sampling to obtain the neutron flux uncertainty contribution from $f_\alpha$.

\begin{figure}[tb]
\includegraphics[width=\columnwidth]{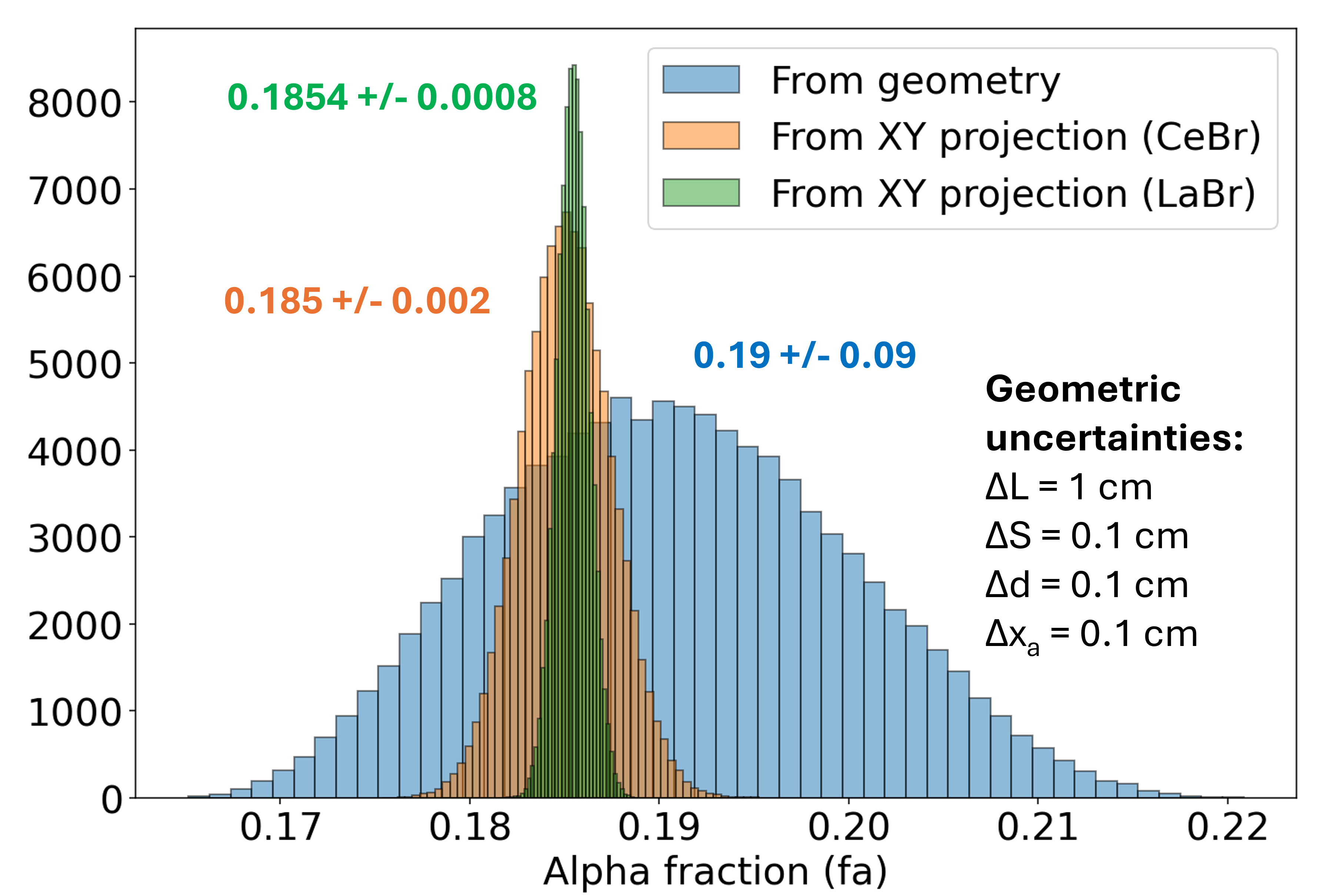}
\caption{\label{fig:fa-dist}Parameter $f_\alpha$ as measured with both LaBr$_3$ and CeBr$_3$, as well as a geometric approximation with reasonable uncertainties.}
\end{figure}

Fig.~\ref{fig:fa-dist} shows the resulting $f_\alpha$ distributions for both LaBr$_3$ and CeBr$_3$, along with an entirely geometric approximation, which is shown to highlight consistency with our model of the setup. The LaBr$_3$ results carry lower uncertainty due to better counting statistics (larger crystal size), which improve the quality of the fit to Eq.~(\ref{eq:supergauss}). Therefore, we used $f_\alpha = 0.1854 \pm 0.0008$ from LaBr$_3$ for the analysis of all the samples.

\begin{figure}[tb]
\includegraphics[width=\columnwidth]{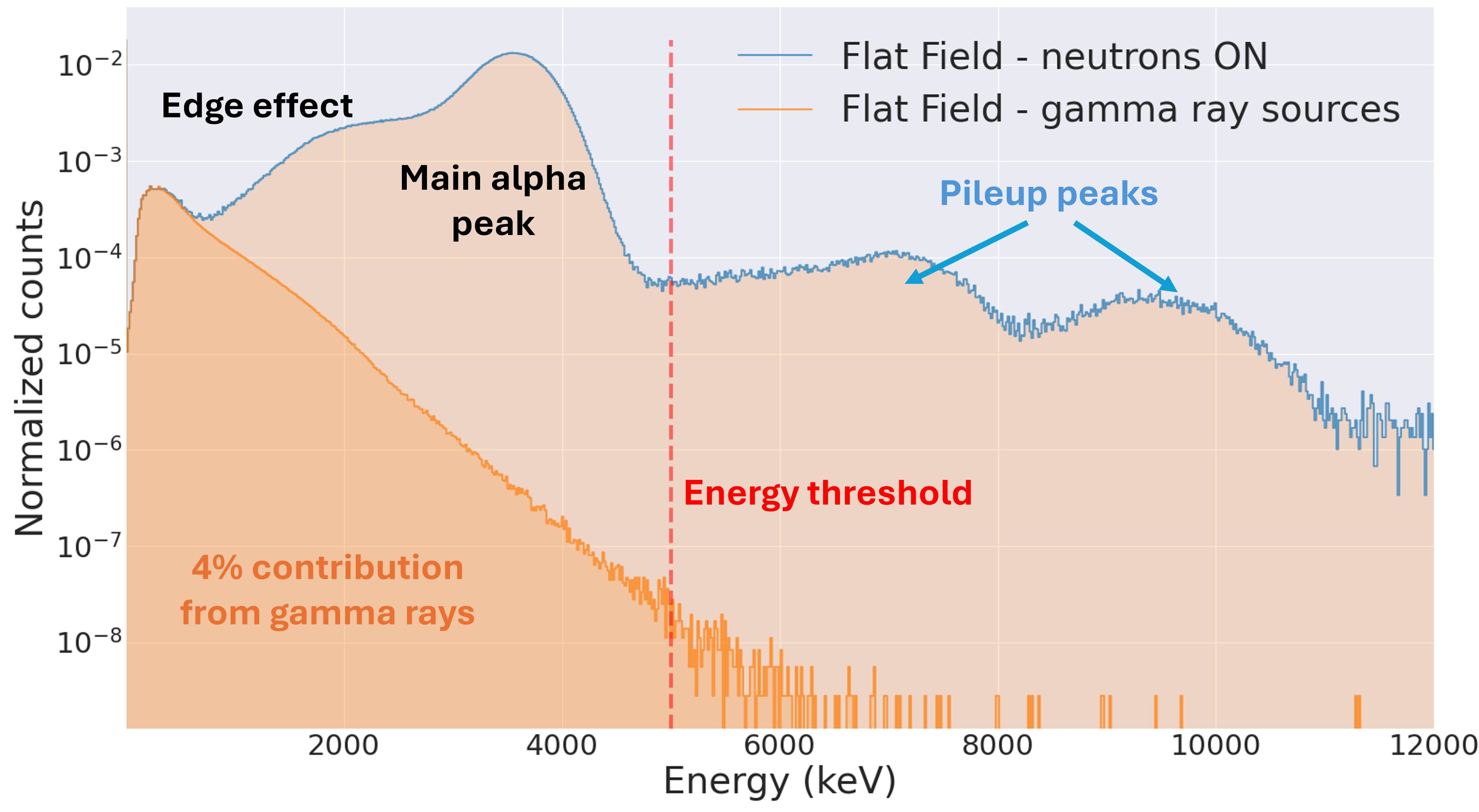}
\caption{\label{fig:alpha-spectrum}Non-coincident (flat field) $\alpha$-particle spectrum and $\gamma$-ray spectrum as measured in the API YAP(Ce) $\alpha$-particle detector.}
\end{figure}

The non-coincident fast alpha triggers ($N_\alpha$) used in Eq.~(\ref{eq:fluence}) requires a small correction, as not all fast triggers in the flat-field measurements originate from 3.5~MeV $\alpha$-particles. Even though the YAP(Ce) is very thin (0.1~mm), $\gamma$-rays will interact and deposit energy in the detector resulting in an exponential distribution as shown in the orange curve in Fig.~\ref{fig:alpha-spectrum}. This $\gamma$-ray contribution to the non-coincident $\alpha$-particle spectrum (flat field) was measured with a variety of $\gamma$-ray sources placed near the $\alpha$-particle detector, and their contribution is approximately 4\% of the total counts. Additionally, there are other charged particles that will produce a signal in the $\alpha$-particle detector, most notably from deuterium-deuterium (DD) and tritium-tritium (TT) reactions:
\begin{eqnarray*}
\text{D} + \text{D} &\rightarrow& {}^3\text{He}\,(0.82\;\text{MeV}) + n\,(2.45\;\text{MeV})\\
\text{D} + \text{D} &\rightarrow& {}^3\text{H}\,(1.01\;\text{MeV}) + p\,(3.02\;\text{MeV}) \\
\text{T} + \text{T} &\rightarrow& \alpha + 2n \quad (Q = 11.3\;\text{MeV})
\end{eqnarray*}

Based on their respective cross sections at 65~kV, as obtained from ENDF/B-VIII.0, the contribution of charged particles other than the 3.5~MeV $\alpha$-particle from DT is approximately 0.86\%. Therefore, the number of fast triggers is reduced by 4.86\% to account for both $\gamma$-rays and other charged particle contributions. We placed a high-energy threshold in the $\alpha$-particle spectrum at approximately 5000~keV to exclude pileup events, as shown in Fig.~\ref{fig:alpha-spectrum}.

Using the $f_\alpha$ results from LaBr$_3$, the corrected $N_\alpha$ from the DAQ, and the known surface area of the targets, the neutron fluence was calculated from Eq.~(\ref{eq:fluence}) for all four irradiated samples. The average neutron flux and measurement times are shown in Table~\ref{tab:flux}.

\begin{table}[b]
\caption{\label{tab:flux}Neutron flux measurements for all irradiated samples.}
\begin{ruledtabular}
\begin{tabular}{lcc}
Sample & Flux (n/cm$^2$/s) & Measurement time (h) \\
\colrule
Fe (1~mm)   & $309 \pm 4$  & 4.0 \\
Fe (8~mm)   & $429 \pm 2$  & 5.0 \\
C (2~mm)    & $446 \pm 5$  & 7.0 \\
C (10~mm)   & $461 \pm 5$  & 6.5 \\
Background  & $455 \pm 5$  & 6.0 \\
\end{tabular}
\end{ruledtabular}
\end{table}

\begin{figure}[tb]
\includegraphics[width=\columnwidth]{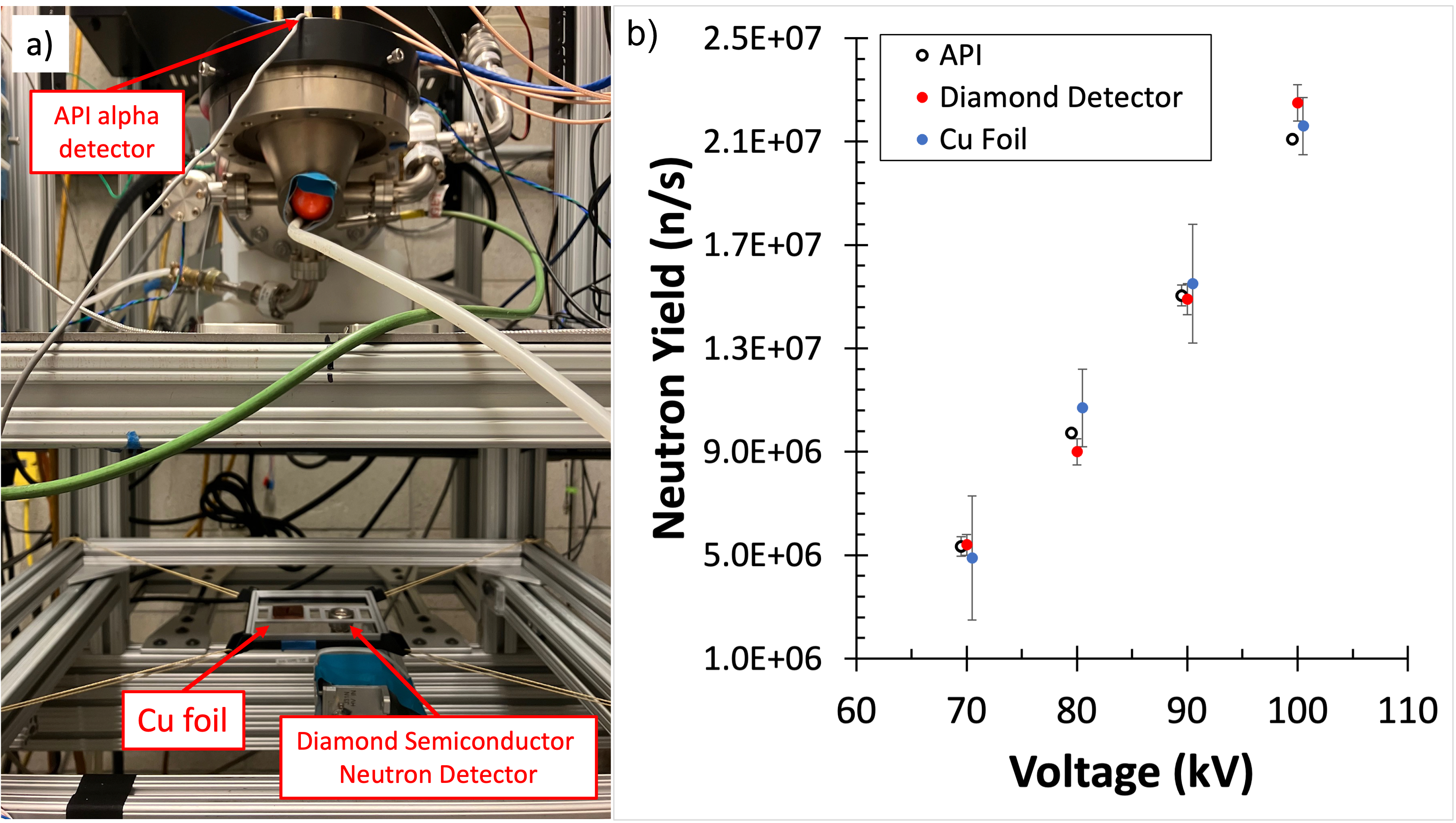}
\caption{\label{fig:flux-independent}(a) Setup used to measure the neutron yield with two independent methods (Cu activation and direct detection with a diamond detector), and (b) results and comparison with API-based neutron flux determination.}
\end{figure}

To further validate the API-based flux determination, we designed two independent cross-checks: (1) irradiation of a standard copper foil using the reaction $^{63}$Cu$(n,2n)^{62}$Cu, followed by offline $\gamma$-ray counting, and (2) direct neutron detection with a co-located diamond detector using the $^{12}$C$(n,\alpha)^{9}$Be reaction. In both cases the witness foil/detector was placed at the sample location and held in place by a low-mass plastic support as shown in Fig.~\ref{fig:flux-independent}a.

The copper foil measurement involved a square target of natural Cu with a purity greater than 99.99\%, side length of 25~mm and 2~mm-thickness. The $^{63}$Cu$(n,2n)^{62}$Cu reaction has an energy threshold of 11.03~MeV, and is therefore relatively insensitive to neutrons downscattered in the room and back into the foil. $^{62}$Cu has a $9.673\pm0.008$~minute half-life to $\beta^+$ decay and the same CeBr$_3$ used in the API system was used to measure the rate of 511~keV emission following $\beta^+$ annihilation. The efficiency was determined by MC simulations. At each beam voltage, the 14~MeV neutron irradiation was carried out for one hour to ensure that secular equilibrium had been reached. Following irradiation, the foil was placed in contact with the end of the CeBr$_3$ detector and the 511~keV peak rate was measured for approximately 5000~s. These measurements were used to infer the activity at the instant that the generator was turned off based on the known half-life. Using the known cross section, foil geometry, nuclear decay data, and the modeled product of efficiency, solid angle and attenuation, we calculated the expected average neutron generator flux at the target position.

The diamond detector included a single crystal with a nominal size of $4.7 \times 4.7 \times 0.5$~mm. These detectors are often used in high flux environments due to their robustness to radiation damage and low efficiency. The response function to 14~MeV neutrons has a characteristic shape including a peak at 8.4~MeV due to the $^{12}$C$(n,\alpha)^{9}$Be reaction. Combining measurements of this peak counting rate with knowledge of the relevant cross section and detector geometry enables the neutron flux to be inferred.

Fig.~\ref{fig:flux-independent}b compares the results of all three methods at different generator voltages. The agreement confirms the reliability of the API-based neutron flux technique.

\subsection{\label{sec:netcounts}Net counts in the photopeak}

\begin{figure}[tb]
\includegraphics[width=\columnwidth]{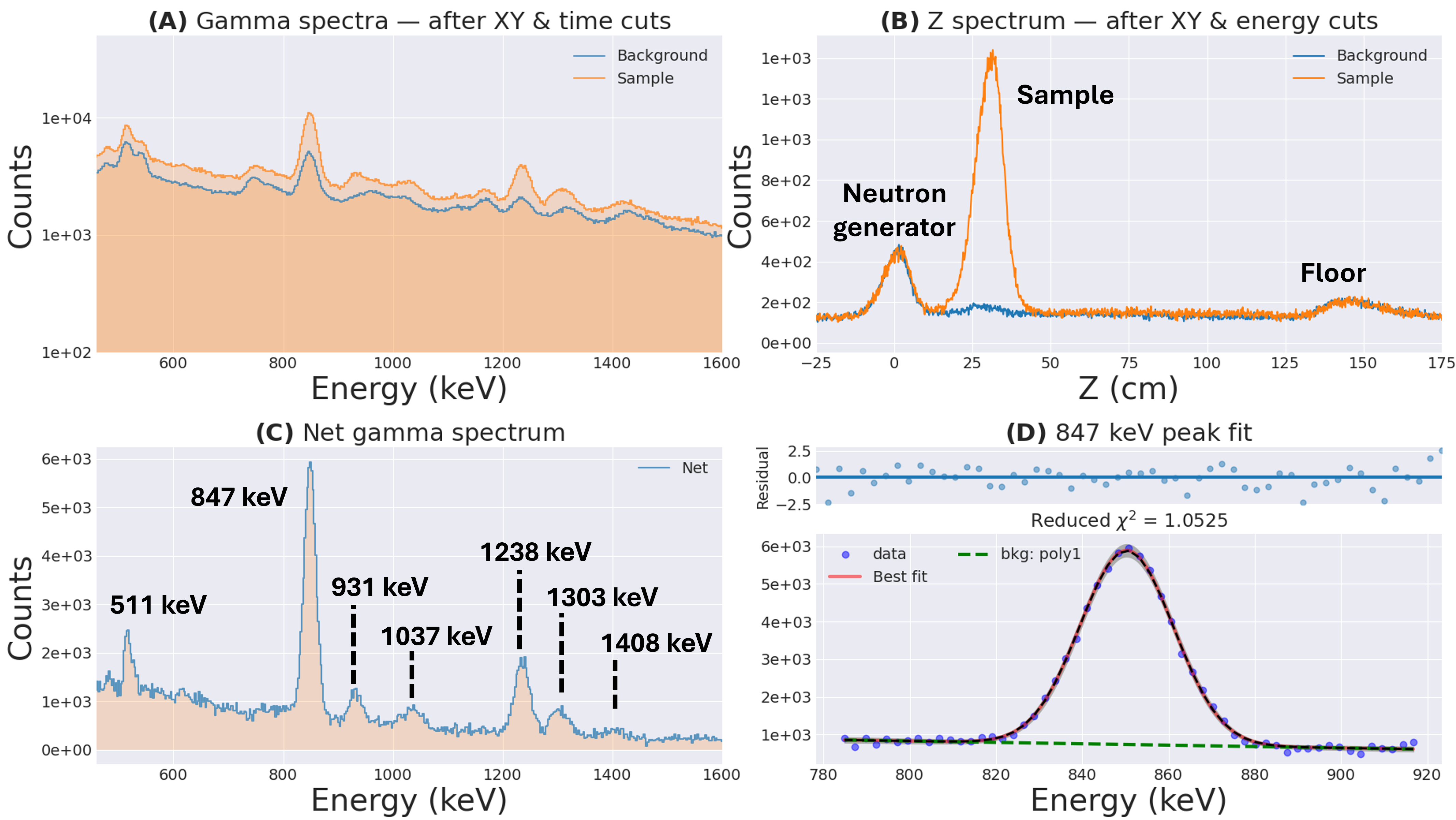}
\caption{\label{fig:netcounts}Prompt API $\gamma$-ray spectra of an 8~mm Fe target, depth ($Z$) distribution, and Gaussian peak fitting with a linear background.}
\end{figure}

The API imaging capabilities were used to substantially suppress room background by applying $x$, $y$, and $z$ cuts during post-processing. Fig.~\ref{fig:netcounts} illustrates this procedure for the 8~mm Fe sample measured with LaBr$_3$, though the same approach was applied to all samples. Fig.~\ref{fig:netcounts}a shows the prompt $\gamma$-ray spectra for the sample and the corresponding background measurement. Fig.~\ref{fig:netcounts}b shows the $Z$ distribution derived from the alpha-gamma time difference spectrum, where distinct structures are visible corresponding to the neutron generator, the sample, and the floor. These peaks are observed at the expected positions from the neutron TOF for 14~MeV neutrons and the experiment geometry. A window of 20--30~cm centered on the sample peak, combined with an $x$-$y$ spatial cut around the sample, substantially reduced the room background. Fig.~\ref{fig:netcounts}c shows the resulting net $\gamma$-ray spectrum with the most prominent peaks identified, and Fig.~\ref{fig:netcounts}d shows the 847~keV peak with its best-fit curve used to extract the net photopeak counts ($N_\gamma$). A summary of net count rates in selected photopeaks across all samples is provided in Table~\ref{tab:xsec}.

Since natural isotopic abundances are used, the measured cross sections cannot be uniquely attributed to a single isotope, with the exception of carbon, for which $^{12}$C$(n,n')^{12}$C is the only reaction producing 4439~keV $\gamma$-rays. For iron, the isotopic composition ($^{54}$Fe: 5.845\%, $^{56}$Fe: 91.754\%, $^{57}$Fe: 2.119\%, $^{58}$Fe: 0.282\%) means that $(n,2n)$ reactions on $^{57}$Fe can also contribute to both the 847~keV and 1238~keV lines. However, this contribution is much smaller compared to the primary $(n,n')$ reactions on $^{56}$Fe. The observed $\gamma$-ray transitions and their most likely production mechanisms are listed in Table~\ref{tab:transitions}.

\begin{table}[tb]
\caption{\label{tab:transitions}Prompt $\gamma$-rays identified in the Fe and C targets and their most likely production mechanisms.}
\begin{ruledtabular}
\begin{tabular}{ll}
$\gamma$-ray energy (keV) & Reactions \\
\colrule
846.78   & $^{56}$Fe$(n,n')^{56}$Fe; $^{57}$Fe$(n,2n)^{56}$Fe \\
931.29   & $^{56}$Fe$(n,2n)^{55}$Fe \\
1037.87  & $^{56}$Fe$(n,n')^{56}$Fe; $^{57}$Fe$(n,2n)^{56}$Fe \\
1238.33  & $^{56}$Fe$(n,n')^{56}$Fe; $^{57}$Fe$(n,2n)^{56}$Fe \\
1303.45  & $^{56}$Fe$(n,n')^{56}$Fe; $^{57}$Fe$(n,2n)^{56}$Fe \\
1408.19  & $^{54}$Fe$(n,n')^{54}$Fe \\
4438.91  & $^{12}$C$(n,n')^{12}$C \\
\end{tabular}
\end{ruledtabular}
\end{table}

\subsection{\label{sec:det-eff}Detector photopeak efficiency}

\begin{figure}[tb]
\includegraphics[width=\columnwidth]{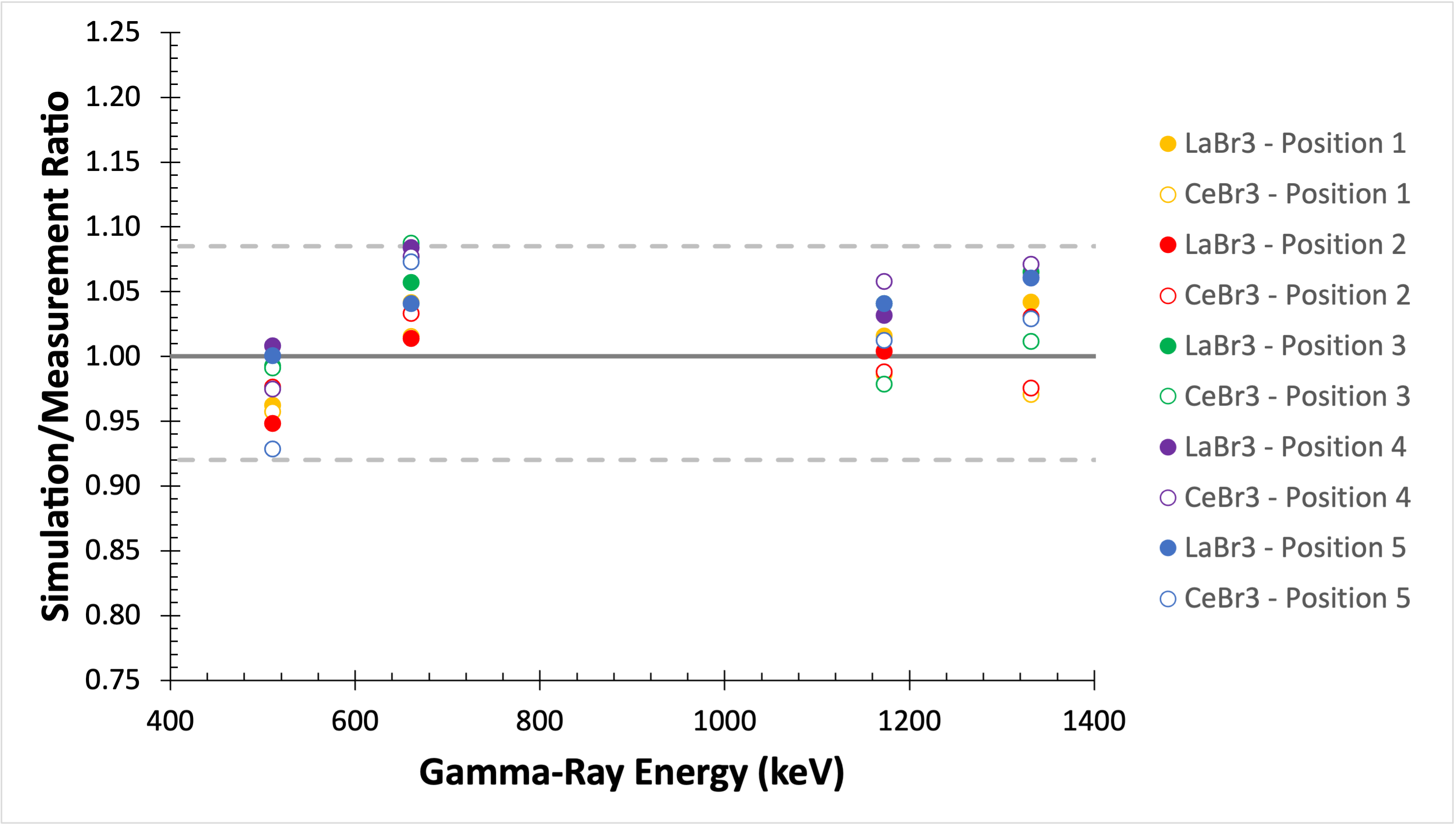}
\caption{\label{fig:efficiency}Comparison of benchmarking measurements, made with $\gamma$-ray check sources, to corresponding GEANT4 simulations, highlighting the precision of the GEANT4 model. All of the data fall within a $\pm 8.5\%$ uncertainty band (dashed lines), which accounts for the systematic uncertainties for the detector array and calibration sources.}
\end{figure}

Both LaBr$_3$ and CeBr$_3$ detectors were characterized in terms of absolute photopeak detection efficiency using check sources of Cs-137, Co-60, Na-22, and Th-228. The sources were placed on a fixture wherein the positions (1 through 5) are located at the center and 4-corner edges of the sample position (see right hand side of Fig.~\ref{fig:setup}). Thus, this efficiency benchmark was performed in the same geometry as the neutron-irradiation sample measurements. Fig.~\ref{fig:efficiency} shows a comparison between the measured photopeak rate for each source and position, divided by the GEANT4-simulated photopeak rate. A value of 1 corresponds to perfect agreement. All measurements fall within a $\pm 8.5\%$ uncertainty band (dashed lines), which accounts for the uncertainties on the source activity (3\%) and detector position (8\%), added in quadrature. This indicates that our quoted systematic uncertainties are an accurate assessment of the uncertainties on $\epsilon  F$.

\subsection{\label{sec:xsec}Prompt gamma production cross sections}

Table~\ref{tab:xsec} presents the measured $\gamma$-ray production cross sections together with the key experimental quantities entering Eq.~(\ref{eq:sigma-rate}), or Eq.~(\ref{eq:sigma}) when multiplied by measurement times. Fig.~\ref{fig:carbon-xsec} and~\ref{fig:iron-xsec} show these results alongside values produced from ENDF/B-VIII.1 and a comparison to other published experimental data.

\begin{table*}
\caption{\label{tab:xsec}Measured $\gamma$-ray production cross sections for Fe and C samples.}
\begin{ruledtabular}
\begin{tabular}{lclclcc}
Sample & Mass (g) & Angle & $\gamma$-transition (keV) & Net count rate (cts/s) & Efficiency $\pm 8.5\%$ & Cross section (mb) \\
\colrule
Fe (1~mm)  & 74.20  & $110^\circ \pm 7^\circ$ (LaBr$_3$) & 846.78  & $0.554 \pm 0.023$   & $3.39\times10^{-3}$ & $727 \pm 71$ \\
Fe (1~mm)  & 74.20  & $110^\circ \pm 7^\circ$ (LaBr$_3$) & 1238.33 & $0.190 \pm 0.020$   & $2.77\times10^{-3}$ & $304 \pm 42$ \\
Fe (1~mm)  & 74.20  & $48^\circ \pm 5^\circ$ (CeBr$_3$)  & 846.78  & $0.0887 \pm 0.0096$ & $5.19\times10^{-4}$ & $759 \pm 106$ \\
Fe (1~mm)  & 74.20  & $48^\circ \pm 5^\circ$ (CeBr$_3$)  & 1238.33 & $0.0287 \pm 0.0131$ & $3.96\times10^{-4}$ & $321 \pm 151$ \\
Fe (8~mm)  & 370.47 & $110^\circ \pm 7^\circ$ (LaBr$_3$) & 846.78  & $3.88 \pm 0.041$    & $2.26\times10^{-3}$ & $676 \pm 60$ \\
Fe (8~mm)  & 370.47 & $110^\circ \pm 7^\circ$ (LaBr$_3$) & 1238.33 & $1.419 \pm 0.037$   & $1.89\times10^{-3}$ & $296 \pm 27$ \\
Fe (8~mm)  & 370.47 & $48^\circ \pm 5^\circ$ (CeBr$_3$)  & 846.78  & $0.817 \pm 0.015$   & $4.345\times10^{-4}$ & $740 \pm 67$ \\
Fe (8~mm)  & 370.47 & $48^\circ \pm 5^\circ$ (CeBr$_3$)  & 1238.33 & $0.281 \pm 0.014$   & $3.24\times10^{-4}$ & $342 \pm 34$ \\
C (2~mm)   & 37.01  & $110^\circ \pm 7^\circ$ (LaBr$_3$) & 4438.91 & $0.125 \pm 0.009$   & $1.09\times10^{-3}$ & $141 \pm 16$ \\
C (2~mm)   & 37.01  & $48^\circ \pm 5^\circ$ (CeBr$_3$)  & 4438.91 & $0.0179 \pm 0.0067$ & $7.88\times10^{-5}$ & $279 \pm 105$ \\
C (10~mm)  & 181.71 & $110^\circ \pm 7^\circ$ (LaBr$_3$) & 4438.91 & $0.615 \pm 0.017$   & $1.02\times10^{-3}$ & $146 \pm 13$ \\
C (10~mm)  & 181.71 & $48^\circ \pm 5^\circ$ (CeBr$_3$)  & 4438.91 & $0.0910 \pm 0.0081$ & $7.78\times10^{-5}$ & $284 \pm 35$ \\
\end{tabular}
\end{ruledtabular}
\end{table*}

\section{\label{sec:discussion}Discussion}

\subsection{\label{sec:uncertainties}A note on uncertainties}

The uncertainty in our measured cross sections is dominated by two factors: counting statistics, which is the primary limitation for thin samples, and the systematic uncertainties associated with the detector efficiency term in Eq.~(\ref{eq:sigma-rate}). The former can be improved by extending measurement times or by increasing the tagged neutron rate through faster $\alpha$-particle detector readout electronics. The latter can be substantially reduced by using calibrated check sources with lower stated activity uncertainties and with a higher-fidelity geometric model of the experiment. The total cross section uncertainty was derived by constructing probability distributions for each factor in Eq.~(\ref{eq:sigma-rate}) and propagating them through a Monte Carlo sampling procedure. Correlations between uncertainty terms were assumed to be negligible; this assumption will be revisited in future work.

\subsection{\label{sec:comparison}Comparison with existing data and Evaluated Nuclear Data Files}

\begin{figure}[tb]
\includegraphics[width=\columnwidth]{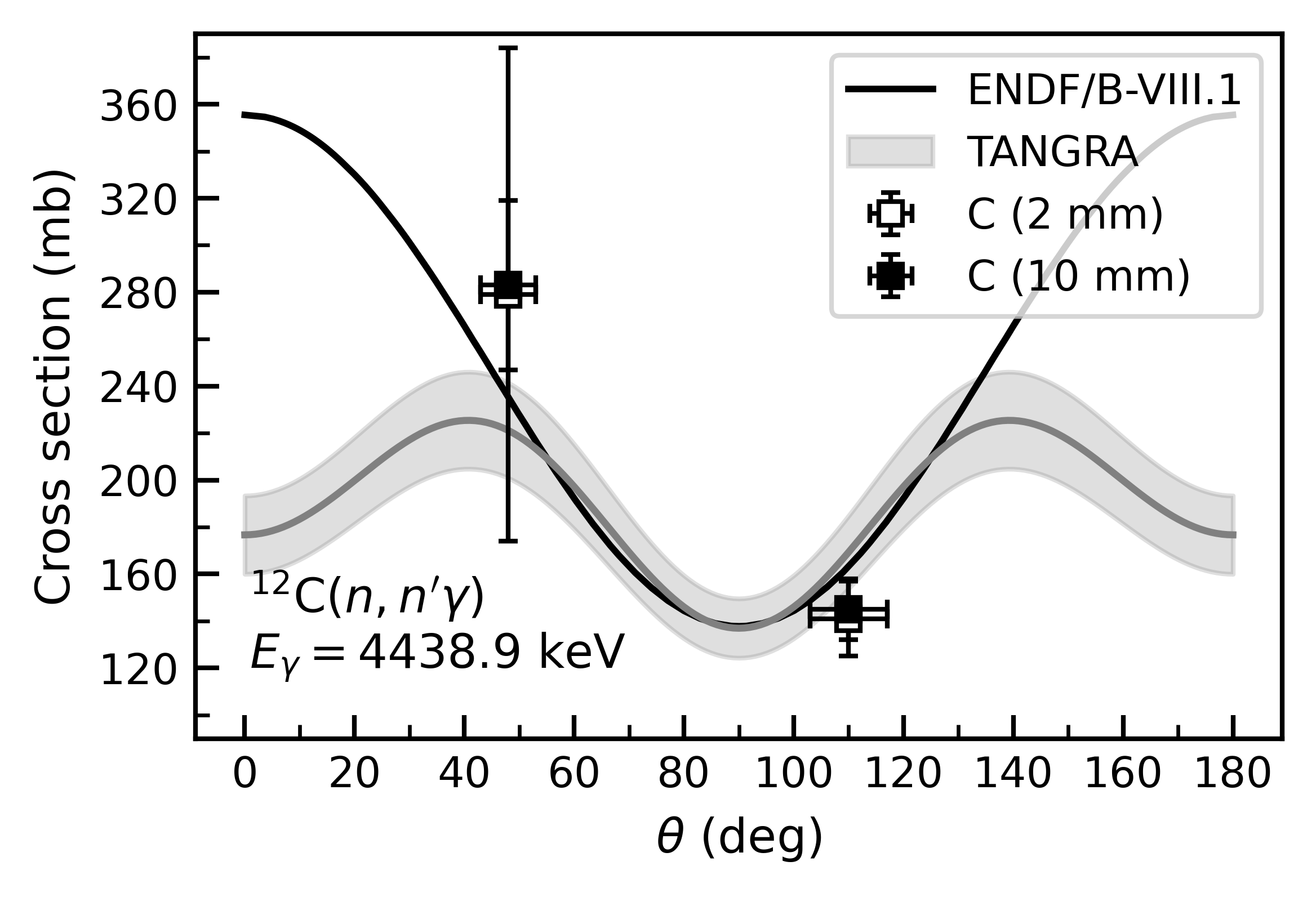}
\caption{\label{fig:carbon-xsec}Angular distribution of the 4438.9-keV $\gamma$-ray production cross section for $^{12}$C induced by 14~MeV neutrons. Square symbols show our measured cross sections for carbon targets of different thicknesses. The solid and shaded curves represent ENDF/B-VIII.1~\cite{Nobre2026} and TANGRA~\cite{Prusachenko2025a}.}
\end{figure}

Fig.~\ref{fig:carbon-xsec} shows good agreement between the present measurements and the ENDF/B-VIII.1 evaluation~\cite{Nobre2026} for the angular distribution of the 4438.9~keV $\gamma$-ray transition from $^{12}$C. This $\gamma$-ray is associated with the decay of the first excited state of $^{12}$C, and can be obtained directly from the population of the discrete inelastic channel MT=51 (in the ENDF-6 format language). We note that higher-lying states in $^{12}$C are above the particle-emission threshold and decay predominantly by alpha emission, so they do not significantly feed the 4438.9~keV transition. We interpret the $\gamma$-ray production as being dominated by direct inelastic scattering to a single state, preserving the angular anisotropy represented in the evaluated data. Within the angular range covered by the present work, the present data are consistent with both the evaluated angular distribution and the TANGRA experimental trend within the experimental uncertainties.

\begin{figure}[tb]
\includegraphics[width=\columnwidth]{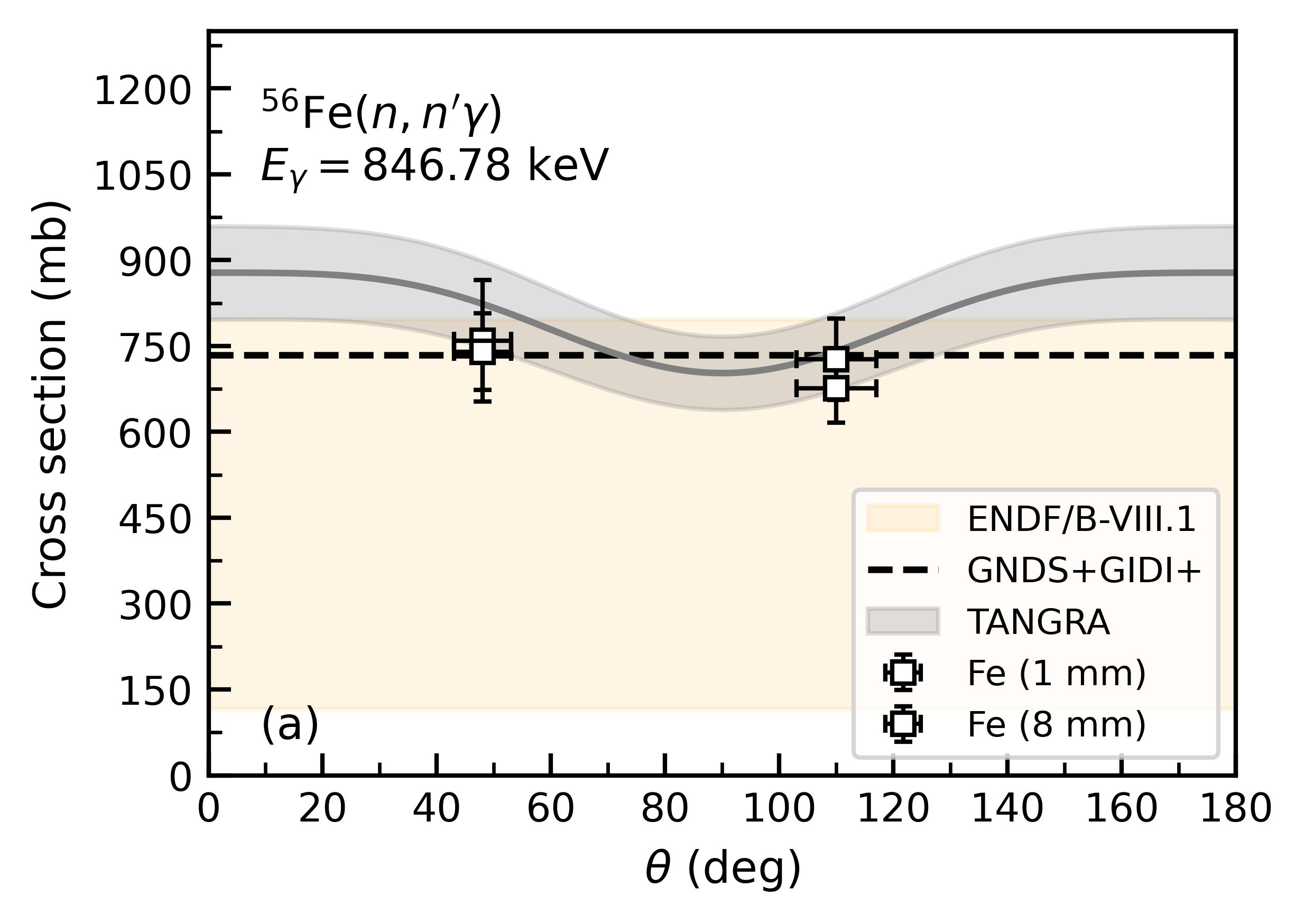}
\caption{\label{fig:iron-xsec}Production cross sections for the 846.78-keV and 1238.33-keV lines from the Fe target induced by 14~MeV neutrons. Square symbols show our measured cross sections for different target thicknesses. Dashed lines show the values obtained with the GNDS+GIDI+ capability from Ref.~\cite{Chimanski2026}. The shaded ENDF/B-VIII.1 bands indicate lower and upper bounds: the lower bound includes direct population and feeding from resolved discrete levels, while the upper bound additionally assumes that the continuum contribution (MT=91) fully feeds the selected transition. TANGRA values are shown as shaded curves~\cite{Prusachenko2025a}.}
\end{figure}

Fig.~\ref{fig:iron-xsec} presents our measurements for the 846.78 and 1238.33-keV transitions in the Fe case. The cross sections measured with both target thicknesses are consistent for both $\gamma$-ray lines. Unlike $^{12}$C, several discrete inelastic channels are open for $^{56}$Fe at this incident energy, together with the unresolved continuum contribution (represented by MT=91 in the ENDF-6 format). As a result, the individual $(n,n'\gamma)$ production cross section cannot be obtained uniquely from the standard ENDF-6 file, because the continuum-to-discrete cascade feeding is not explicitly available in the file. In this case we limit ourselves to show conservative band limits obtained directly from the ENDF-6 information. The lower limit includes the direct population of the emitting level plus feeding from higher resolved discrete states. The upper limit adds the full MT=91 contribution, which is equivalent to assuming that the entire continuum feeds the selected transition.

We also show the result obtained with the recently presented method in Ref.~\cite{Chimanski2026}, referred to as GNDS+GIDI+, which uses an extended version of the evaluated file in the GNDS format. In this framework, the gamma cascade can be reconstructed explicitly, including continuum-to-continuum and continuum-to-discrete transitions. The $\gamma$-ray production cross section was obtained by simulating a large number of cascade events, computing the emission probability of a given $\gamma$-ray, and multiplying this probability by the inelastic cross section from the evaluated file. The GNDS+GIDI+ results fall within the ENDF/B-VIII.1 bounds and are in good agreement with our measurements. The weak angular dependence observed for $^{56}$Fe is also consistent with the reaction strength being distributed over many higher-lying levels and cascade paths, which tend to reduce the alignment of the emitting state.

\subsection{\label{sec:neutron-energy}Neutron interaction energy}

\begin{figure}[tb]
\includegraphics[width=\columnwidth]{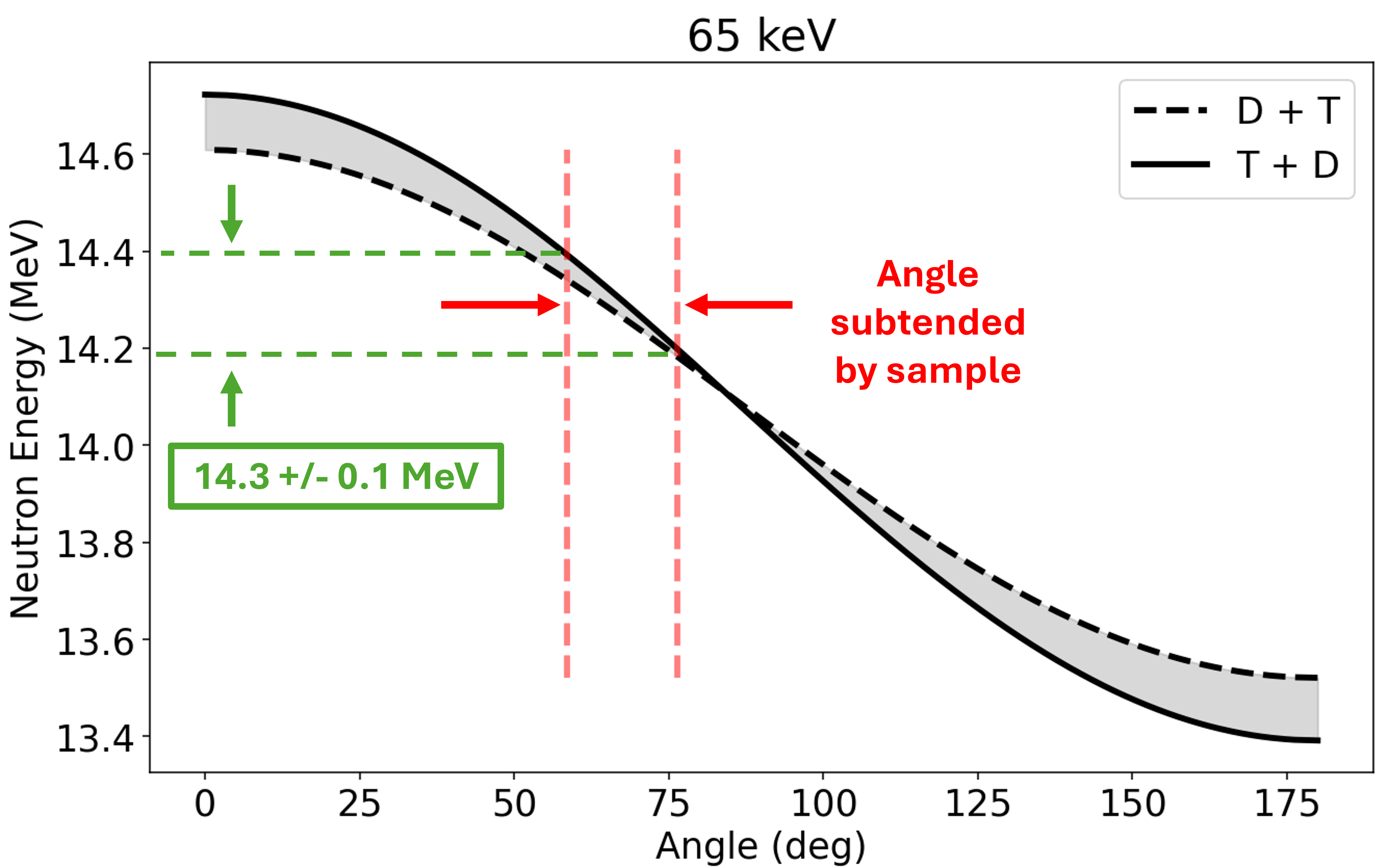}
\caption{\label{fig:neutron-energy}Neutron energy distribution due to a deuterium or tritium beam accelerated at 65~kV. In red is the angle subtended by the square sample, and in green is the maximum expected neutron energy spread.}
\end{figure}

The analysis assumes a nominal neutron interaction energy of 14.1~MeV. In practice, however, the DT reaction produces a narrow distribution of neutron energies centered on that value, arising from the kinematics of the reaction and the beam ion species present. Because we operate a mixed DT beam, both D(T,$n$)$\alpha$ and T(D,$n$)$\alpha$ reactions occur with equal probability, and additional ion species (D$_2$, T$_2$, etc.) contribute a spread of energies that falls between the two monatomic cases. Furthermore, the neutron generator is tilted $22.5^\circ$ from horizontal (see the Adelphi DT108 design in Ref.~\cite{Adelphi}), so the sample does not sit at $90^\circ$ relative to the beam axis. Accounting for the solid angle subtended by the sample at 32~cm from the beam spot and edge cases in the reaction kinematics, the neutron energy distribution on the sample is $14.3 \pm 0.1$~MeV, as shown in Fig.~\ref{fig:neutron-energy}.

\subsection{\label{sec:multiple-scattering}Effects of multiple neutron scattering in the sample}

\begin{figure}[tb]
\includegraphics[width=\columnwidth]{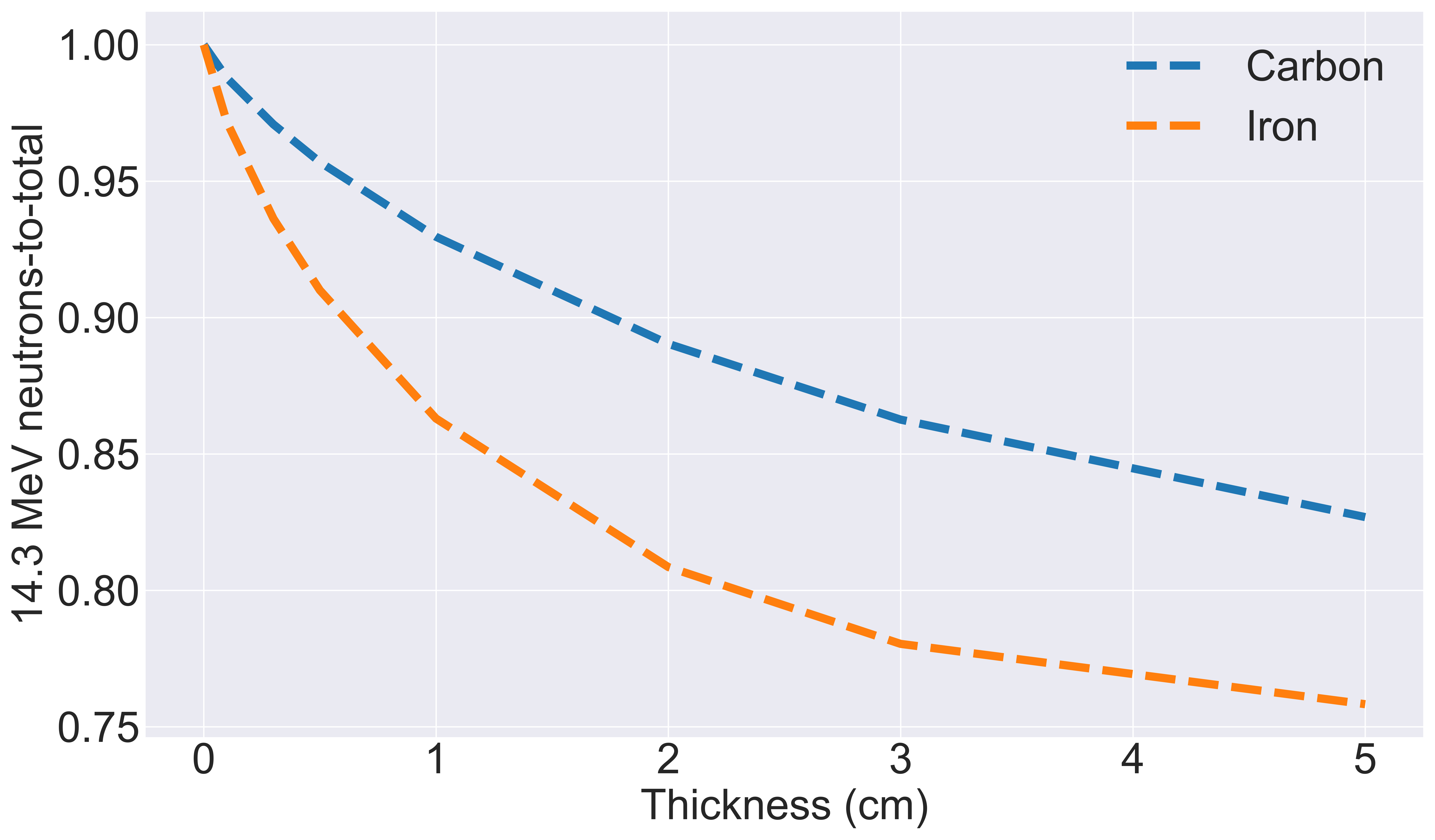}
\caption{\label{fig:n-scattering}MCNP simulations of 14.3~MeV neutron scattering in both samples of iron and carbon. The $y$-axis indicates the ratio of unscattered 14.3~MeV neutrons to total neutrons as they go through the samples.}
\end{figure}

An additional consideration is multiple neutron scattering within the sample, which can produce $\gamma$-rays at neutron energies below 14.3~MeV. For example, a 14.3~MeV primary neutron may scatter off a nucleus near the sample surface, lose energy, and then inelastically scatter off a second nucleus, whose de-excitation $\gamma$-ray is subsequently detected in coincidence with the $\alpha$-particle. To quantify this effect, we performed MCNP simulations tallying neutron energies within the sample inside a 3~ns time window (Fig.~\ref{fig:n-scattering}). For the 1~mm Fe sample, more than 97\% of reactions occur at 14.3~MeV; for the 2~mm C sample, this fraction exceeds 98\%. The thin-sample cross sections therefore predominantly reflect single-scatter interactions with 14.3~MeV neutrons. For the thicker samples, the uncollided fraction drops to approximately 88\% for 8~mm Fe and 93\% for 10~mm C, meaning a larger share of collided primary neutrons contributes to those measurements. Both thin and thick sample measurements are nonetheless valuable: thin samples yield differential cross sections at 14.3~MeV, while thick samples probe integral quantities needed for benchmarking radiation transport codes.

\section{\label{sec:summary}Summary and outlook}

We have presented a compact, laboratory-scale technique for precision $\gamma$-ray production cross section measurements at an incident neutron energy of 14.3~MeV. Fe and C were selected as benchmark samples due to the availability of extensive published data for comparison across multiple experimental techniques. The results obtained using API show good agreement with the literature while offering significant advantages. In particular, the use of API substantially reduces systematic uncertainties in both the neutron flux determination and the geometric setup, which are typically the dominant limitations in conventional methods. This technique has the potential to address gaps and discrepancies in existing cross-section libraries at a fraction of the cost of large-scale dedicated facilities, with direct benefits for active neutron interrogation systems, detector calibration activities, and nuclear fusion research (including materials testing, diagnostics, and activation studies). We are currently extending these measurements to a broader set of high-purity elemental samples to build a comprehensive database, the Berkeley Atlas, analogous to the Baghdad Atlas~\cite{Hurst2021}, which remains one of the best experimental references for $\gamma$-ray production cross sections from neutron inelastic scattering in a fission reactor fast neutron spectrum.

To improve upon the current measurements, we are integrating HPGe detectors with the API system, which will provide higher energy resolution, better signal-to-noise ratio, and reduced ambiguity from overlapping $\gamma$-ray lines. The primary technical challenges are achieving fast timing with HPGe and managing high neutron rates.

Finally, future work will explore triple-coincidence measurements, in which the outgoing neutron is detected in coincidence with both the $\alpha$-particle and the prompt $\gamma$-ray. This additional constraint on the reaction kinematics will enable higher-fidelity cross section evaluations and provide deeper insight into nuclear structure. In addition, we plan to develop GEANT4 simulations of our setup with integrated GNDS/GIDI capabilities~\cite{Chimanski2026} to provide a direct benchmark of evaluated nuclear data within the experimental framework.

\begin{acknowledgments}
The information, data, or work presented herein was funded by:
An appointment to the NASA Postdoctoral Program (NPP) at the NASA Goddard Space Flight Center, administered by Oak Ridge Associated Universities under contract with NASA,
The Internal Scientist Funding Model (ISFM) through the Fundamental Laboratory Research (FLaRe) program at NASA Goddard Space Flight Center, and
The Nuclear Data Inter Agency Working Group (NDIAWG) Research Program, U.S. Department of Energy, under Contract No. DE-AC02-05CH11231.
\end{acknowledgments}

\section*{Data availability}
The data and processing code supporting the findings of this article are openly available in~\cite{ayllon_unzueta_2026_zenodo}.

\bibliography{FLARE-APS}

\end{document}